\documentclass[aps,prd,twocolumn,nofootinbib,superscriptaddress,floatfix]{revtex4-2}

\usepackage[T1]{fontenc}
\usepackage[utf8]{inputenc}
\usepackage{amsmath,amssymb,amsfonts,bm}
\usepackage{graphicx}
\usepackage{booktabs}
\usepackage{hyperref}
\usepackage{xcolor}

\hypersetup{
  colorlinks=true,
  linkcolor=blue,
  citecolor=blue,
  urlcolor=blue
}

\newcommand{\ket}[1]{\left|#1\right\rangle}
\newcommand{\bra}[1]{\left\langle#1\right|}

\newcommand{\ident}{\mathbb{I}}
\newcommand{\Tr}{\mathrm{Tr}}
\newcommand{\dd}{\mathrm{d}}

\begin{document}

\title{Internal pseudospin, lepton-number superselection, and neutrino--antineutrino coherence in massive neutral-fermion one-particle states}

\author{R. Romero}
\affiliation{Departamento de Ciencias Naturales, Unidad Cuajimalpa, Universidad Aut\'onoma Metropolitana, Av. Vasco de Quiroga 4871, Santa Fe Cuajimalpa, Ciudad de M\'exico 05348, M\'exico}

\date{\today}

\begin{abstract}
At fixed three-momentum, massive Dirac neutrino one-particle states span a 4D space of particle--antiparticle identity and helicity. We show that helicity flip, charge conjugation, and their product close an internal $SU(2)$ pseudospin subalgebra within $SU(4)$, distinct from the Wigner little group. Its helicity generator is the lepton-number-weighted spin rotation $U_1=2LJ_2$. The lepton-number $L$ and helicity $\mathcal H_h$ grade the 16 generators $\tau_\mu\otimes\sigma_\nu$, organizing the $\Delta L=2$ sector. Helicity-preserving directions $\tau_{1,2}\otimes\sigma_{0,3}$ carry the pseudo-Dirac mixing $\kappa_{\rm pD}=\Delta m^2/4E$ (active--sterile), while helicity-flipping directions $\tau_{1,2}\otimes\sigma_{1,2}$ carry the neutrinoless double-beta decay mass factor (active--active). Furthermore, charge conjugation matches the $U_2$ generator. The Majorana condition is thus a projection onto the $U_2=+1$ eigenspace, where only the Wigner algebra survives. This framework algebraically classifies Majorana masses and pseudo-Dirac splittings without assuming neutrinos are Majorana particles.
\end{abstract}

\maketitle

\section{Introduction}

The distinction between Dirac and Majorana neutrinos is ultimately a statement about the physical status of lepton number.  A Dirac neutrino carries an independent conserved charge, whereas a Majorana neutrino is a neutral fermion whose particle and antiparticle labels are not independent degrees of freedom \cite{Majorana1937,GiuntiKim2007,MohapatraPal2004}.  Oscillation experiments establish that neutrinos are massive, but they do not determine whether the mass eigenstates carry an exact conserved $U(1)_L$ charge.  The canonical laboratory probe is neutrinoless double-beta decay, whose observation would imply lepton-number violation and, under the standard light-neutrino exchange interpretation, Majorana neutrino masses \cite{SchechterValle1982,BilenkyGiunti2015}.  Conversely, the absence of a signal cannot prove that neutrinos are Dirac particles \cite{Hirsch2018}.

The practical difficulty is that many differences between Dirac and Majorana neutrinos are suppressed for ultra-relativistic particles.  In a broad class of amplitudes, the distinction is hidden by powers of $m_\nu/E$ or by the inability to measure the final helicity and lepton-number assignment \cite{LiWilczek1982,KayserShrock1982}.  This is the content of the practical Dirac--Majorana confusion theorem and its modern variants \cite{Rodejohann2017,Kim2023}.  A related lesson from ordinary neutrino oscillation theory is that coherent superpositions of mass eigenstates are physical because the interfering components carry the same conserved charges and can be produced and detected coherently \cite{Kayser1981,GiuntiKimLee1991,GiuntiKimLee1998}.  This is distinct from coherence between states with different values of an exactly superselected charge.

A subtle representation-theoretic version of the same issue appears in the one-particle space of a massive neutral Dirac field.  For a massive spin-$1/2$ particle, the Wigner little group is $SU(2)$ and acts on spin labels inside a fixed irreducible representation \cite{Wigner1939,Weinberg1964,WeinbergQFT1}.  However, a neutral Dirac field contains, at fixed momentum, four free one-particle states: two helicities for the particle and two helicities for the antiparticle.  A previous construction exhibited three unitary transformations on this four-state space: a helicity flip, charge conjugation, and their product, the last of which maps a left-helicity neutrino state to a right-helicity antineutrino state up to phases \cite{Romero2021}.  These transformations commute with the free Hamiltonian and momentum and close an $SU(2)$ algebra.

The purpose of the present work is to clarify the physical meaning of this algebra and to connect it to explicit one-particle Hamiltonian terms.  The guiding point is conservative: the existence of a unitary map inside a degenerate free state space does not automatically imply a new physical particle identity or a dynamical transition.  A parallel example is provided by charge-conjugation eigenspinors.  Algebraic charge-conjugation properties of a spinor basis do not by themselves define a new fermion species or a new field equation; for instance, c-number Elko spinors can be unitarily related to massless Weyl bispinors and obey the massless Dirac equation \cite{Romero2023}.  In the neutrino case, the corresponding distinction is between a kinematical algebra in an enlarged one-particle space and the dynamics allowed by lepton-number conservation or violation.  This manuscript is intended as a formal classification and does not assume any particular microscopic model for $\Delta L=2$ dynamics.

We show that the three transformations of Ref.~\cite{Romero2021} form an internal pseudospin algebra embedded in the $SU(4)$ algebra of the fixed-momentum space.  They are little-group-like only in the limited sense that they preserve free four-momentum.  They are not the ordinary Wigner spin little group because two generators exchange particle and antiparticle sectors.  The physical status of the algebra is controlled by the lepton-number operator $L$: one generator commutes with $L$, while two anticommute with it.  Exact lepton-number superselection therefore makes the charge-conjugating directions kinematically well-defined but dynamically inactive.  If $U(1)_L$ is broken, those same directions are the natural channels for neutrino--antineutrino coherence.

The second aim is to make this classification operational.  We introduce a tensor basis $\tau_\mu\otimes\sigma_\nu$, where $\tau_i$ acts on the particle--antiparticle label and $\sigma_i$ on the helicity label.  Lepton number and the helicity label are commuting involutions, and they grade this basis into four classes.  The grading is the organizing device of the paper: it identifies the three generators as one representative per nontrivial class, and it separates the lepton-number-violating sector into a helicity-preserving part, which carries the pseudo-Dirac oscillation and is not amplitude-suppressed, and a helicity-flipping part, which carries neutrino--antineutrino coherence and is suppressed by the neutrino mass.  We give explicit Dirac, Majorana, and pseudo-Dirac limits, show that charge conjugation is one of the three generators and that the Majorana condition is the corresponding projection, derive the pseudo-Dirac one-particle Hamiltonian from the Weyl mass matrix, and reduce each $\Delta L=2$ class to a minimal two-state model.

\section{Fixed-momentum one-particle space}
\label{sec:space}

Consider a massive neutral Dirac field and restrict attention to the free one-particle subspace at fixed three-momentum $\mathbf p$.  We order the basis as
\begin{equation}
\mathcal B_{\mathbf p}=\left(\ket{\nu_-},\ket{\nu_+},\ket{\bar\nu_-},\ket{\bar\nu_+}\right),
\label{eq:basis}
\end{equation}
where the subscripts denote helicity.  The common momentum label is suppressed.  In the free theory,
\begin{equation}
H_0=E_{\mathbf p}\,\ident_4,\qquad E_{\mathbf p}=\sqrt{\mathbf p^2+m^2},
\label{eq:freeH}
\end{equation}
within this fixed-momentum subspace.  The degeneracy in Eq.~\eqref{eq:freeH} is the kinematical origin of the enlarged internal rotation freedom.

It is useful to factor the four-state space as
\begin{equation}
\mathcal H_{\mathbf p}\simeq \mathcal H_{\nu\bar\nu}\otimes\mathcal H_h,
\label{eq:tensor}
\end{equation}
where the first factor is spanned by the particle--antiparticle label $(\nu,\bar\nu)$ and the second by the helicity label $(-,+)$.  Let $\tau_i$ denote Pauli matrices acting on $\mathcal H_{\nu\bar\nu}$ and $\sigma_i$ Pauli matrices acting on $\mathcal H_h$, with $\tau_0=\sigma_0=\ident_2$.  In this convention, the lepton-number operator is
\begin{equation}
L=\tau_3\otimes\ident_2
=\begin{pmatrix}
1&0&0&0\\
0&1&0&0\\
0&0&-1&0\\
0&0&0&-1
\end{pmatrix}.
\label{eq:L}
\end{equation}
The helicity-label operator may be represented as
\begin{equation}
\mathcal H_h=\ident_2\otimes\sigma_3,
\label{eq:helicitylabel}
\end{equation}
with the sign convention set by the basis ordering in Eq.~\eqref{eq:basis}.  A reversal of the helicity ordering merely conjugates the following formulas by a permutation matrix and does not alter the lepton-number classification.

The physical distinction between free degeneracy and exact symmetry is essential.  Since $H_0$ is proportional to the identity on $\mathcal H_{\mathbf p}$, any unitary $V\in U(4)$ obeys
\begin{equation}
[V,H_0]=0.
\label{eq:anyU}
\end{equation}
Equation~\eqref{eq:anyU} is therefore not sufficient to make $V$ a physically realizable symmetry of the full theory.  Interactions and conserved charges select which subalgebras survive as dynamical or observable transformations.

One caveat should be stated at the outset.  If lepton number is exactly conserved and superselected in the sense of Wick, Wightman, and Wigner \cite{WickWightmanWigner1952}, then the coherent superpositions of $L=+1$ and $L=-1$ states contained in $\mathcal H_{\mathbf p}$ are not realizable states.  Throughout this work $\mathcal H_{\mathbf p}$ is used as a formal carrier space on which the algebra acts, not as a Hilbert space of physically preparable one-particle states.  Which of its rays are physical is exactly the question addressed in Sec.~\ref{sec:superselection}.

\section{Embedded $SU(2)$ algebra and comparison with Wigner spin}
\label{sec:su2}

In the basis \eqref{eq:basis}, the three transformations of interest can be written compactly as
\begin{align}
U_1&=\tau_3\otimes\sigma_2,\label{eq:U1}\\
U_2&=\tau_1\otimes\ident_2,\label{eq:U2}\\
U_3&=\tau_2\otimes\sigma_2.\label{eq:U3}
\end{align}
Explicitly,
\begin{align}
U_1&=\begin{pmatrix}
0&-i&0&0\\
i&0&0&0\\
0&0&0&i\\
0&0&-i&0
\end{pmatrix},\label{eq:U1mat}\\
U_2&=\begin{pmatrix}
0&0&1&0\\
0&0&0&1\\
1&0&0&0\\
0&1&0&0
\end{pmatrix},\label{eq:U2mat}\\
U_3&=\begin{pmatrix}
0&0&0&-1\\
0&0&1&0\\
0&1&0&0\\
-1&0&0&0
\end{pmatrix}.\label{eq:U3mat}
\end{align}
These matrices reproduce the state transformations of Ref.~\cite{Romero2021}, up to the ordering and phase convention.  Their action is transparent: $U_1$ flips helicity without changing particle into antiparticle, $U_2$ exchanges particle and antiparticle at fixed helicity, and $U_3$ exchanges particle and antiparticle while reversing helicity.  In particular,
\begin{equation}
U_3\ket{\nu_-}=-\ket{\bar\nu_+},\qquad
U_3\ket{\bar\nu_+}=-\ket{\nu_-},
\label{eq:U3active}
\end{equation}
which identifies the active-helicity channel relevant for Majorana-like neutrino--antineutrino coherence.

\begin{figure}[t]
\centering
\includegraphics[width=0.98\columnwidth]{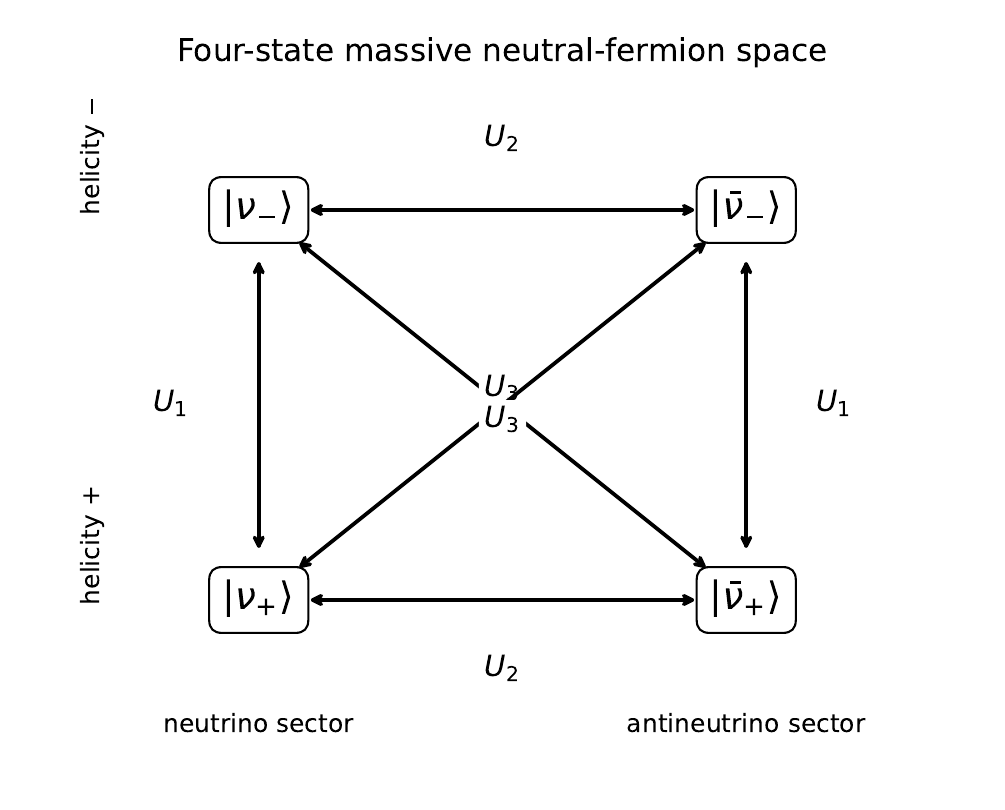}
\caption{Four-state fixed-momentum space of a massive neutral Dirac fermion.  The operator $U_1$ flips helicity within the same particle--antiparticle sector, $U_2$ exchanges particle and antiparticle at fixed helicity, and $U_3$ combines both operations.  The diagonal $U_3$ channel connects $\nu_-$ with $\bar\nu_+$ and $\nu_+$ with $\bar\nu_-$.}
\label{fig:fourstate}
\end{figure}

The matrices are Hermitian and unitary:
\begin{equation}
U_i^\dagger=U_i,
\qquad
U_i^2=\ident_4,
\qquad
U_iU_i^\dagger=\ident_4.
\label{eq:unitaryHermitian}
\end{equation}
Using the Pauli algebra gives
\begin{equation}
[U_i,U_j]=2i\epsilon_{ijk}U_k.
\label{eq:su2comm}
\end{equation}
Thus the normalized generators
\begin{equation}
T_i=\frac{1}{2}U_i
\label{eq:Ti}
\end{equation}
satisfy
\begin{equation}
[T_i,T_j]=i\epsilon_{ijk}T_k.
\label{eq:TiComm}
\end{equation}
The algebra is therefore an $SU(2)$ subalgebra embedded in the $SU(4)$ algebra of the four-dimensional fixed-momentum space:
\begin{equation}
SU(2)_{\nu\bar\nu}\subset SU(4)_{\rm one\mbox{-}particle}.
\label{eq:embedding}
\end{equation}
Figure~\ref{fig:fourstate} summarizes the action of the three generators.

The ordinary Wigner spin generators may be represented, in the same tensor notation and in a fixed spin basis, as
\begin{equation}
J_i=\frac{1}{2}\,\tau_0\otimes\sigma_i.
\label{eq:WignerJ}
\end{equation}
They satisfy the same abstract commutation relations,
\begin{equation}
[J_i,J_j]=i\epsilon_{ijk}J_k,
\label{eq:Jcomm}
\end{equation}
and have the same spin-$1/2$ Casimir on each particle--antiparticle copy,
\begin{equation}
\mathbf J^2=\sum_i J_i^2=\frac{3}{4}\ident_4.
\label{eq:Jcasimir}
\end{equation}
The internal pseudospin generators also obey
\begin{equation}
\mathbf T^2=\sum_i T_i^2=\frac{3}{4}\ident_4.
\label{eq:Tcasimir}
\end{equation}
The equality of Casimirs shows that both algebras act as two copies of a spin-$1/2$ representation on the four-dimensional vector space.  The difference is not the abstract $SU(2)$ representation but its embedding relative to physically conserved labels.  The Wigner spin generators commute with lepton number,
\begin{equation}
[J_i,L]=0,
\label{eq:JL}
\end{equation}
whereas, as shown below, two of the $T_i$ exchange the lepton-number sectors.  This is why preserving free four-momentum is not enough to identify $SU(2)_{\nu\bar\nu}$ with the ordinary massive-particle little group.  The former is an internal pseudospin algebra in a degenerate free space; the latter is the spin little group acting inside each particle sector.

The relation between $U_1$ and ordinary spin can be made sharper.  Since $\sigma_2$ is itself a Hermitian involution, $\tau_0\otimes\sigma_2=2J_2$, and direct computation gives
\begin{equation}
U_1=2\,L\,J_2,
\qquad
[L,J_2]=0.
\label{eq:U1LJ2}
\end{equation}
The helicity-flip generator of $SU(2)_{\nu\bar\nu}$ is therefore not a Wigner spin rotation but the \emph{lepton-number-weighted} one: it acts with opposite sign in the $\nu$ and $\bar\nu$ sectors.  Similarly $U_3=-iU_1U_2$, so the whole triple is generated by $J_2$, $L$, and charge conjugation.  This is the algebraic content of the statement that $SU(2)_{\nu\bar\nu}$ is not the spin little group.

\subsection{A $\mathbb Z_2\times\mathbb Z_2$ grading}
\label{subsec:grading}

The group $SU(4)$ contains many $SU(2)$ subalgebras, and the triple \eqref{eq:U1}--\eqref{eq:U3} is not unique.  It is nevertheless distinguished, for the following reason.  The lepton-number operator $L=\tau_3\otimes\ident_2$ and the helicity label $\mathcal H_h=\ident_2\otimes\sigma_3$ are commuting Hermitian involutions,
\begin{equation}
L^2=\mathcal H_h^2=\ident_4,
\qquad
[L,\mathcal H_h]=0.
\label{eq:involutions}
\end{equation}
Every element of the tensor basis either commutes or anticommutes with each of them.  The sixteen operators $\tau_\mu\otimes\sigma_\nu$ therefore split into four $\mathbb Z_2\times\mathbb Z_2$ grading classes, labeled by $(\Delta L,\Delta h)$:
\begin{align}
(0,0):&\quad \tau_{0,3}\otimes\sigma_{0,3},\label{eq:class00}\\
(0,1):&\quad \tau_{0,3}\otimes\sigma_{1,2}\ \ni\ U_1,\label{eq:class01}\\
(2,0):&\quad \tau_{1,2}\otimes\sigma_{0,3}\ \ni\ U_2,\label{eq:class20}\\
(2,1):&\quad \tau_{1,2}\otimes\sigma_{1,2}\ \ni\ U_3.\label{eq:class21}
\end{align}
Here $\Delta L=2$ means anticommutation with $L$, and $\Delta h=1$ means anticommutation with $\mathcal H_h$, i.e.\ reversal of the helicity label.  The three generators $U_1$, $U_2$, $U_3$ are one representative of each of the three nontrivial classes, and closure of the algebra fixes the remaining freedom up to phase conventions and relabeling of the transverse helicity axes.  This is the precise sense in which the triple is singled out: not by uniqueness inside $SU(4)$, but as one generator per nontrivial $(L,\mathcal H_h)$ grading class.

The product of the two involutions,
\begin{equation}
W\equiv L\,\mathcal H_h=\tau_3\otimes\sigma_3,
\label{eq:Wop}
\end{equation}
has eigenvalue $+1$ on $\{\ket{\nu_-},\ket{\bar\nu_+}\}$ and $-1$ on $\{\ket{\nu_+},\ket{\bar\nu_-}\}$.  As shown in Sec.~\ref{subsec:pseudodirac}, the first pair consists of the states created and annihilated by a left-chiral field and the second of those associated with a right-chiral field.  Thus $W$ is the weak-interaction ``active versus sterile'' label, and it coincides with chirality up to $\mathcal O(m/E)$ corrections.  One finds
\begin{equation}
[U_3,W]=0,
\qquad
\{U_1,W\}=\{U_2,W\}=0,
\label{eq:UW}
\end{equation}
so $U_3$ maps active states to active states, while $U_1$ and $U_2$ map active states to sterile ones.  Section~\ref{sec:helicitySelection} shows that this distinction, rather than lepton number alone, controls which $\Delta L=2$ observable a given tensor sector produces.

\section{Helicity, chirality, and charge conjugation}
\label{sec:helicityC}

The label $h=\pm$ in Eq.~\eqref{eq:basis} denotes helicity.  For a massive particle helicity is not Lorentz invariant: a sufficiently boosted observer can reverse the sign of the momentum relative to the spin.  Chirality, by contrast, is the eigenvalue of $\gamma^5$ and is Lorentz covariant, but for a massive fermion chiral and helicity eigenstates do not coincide exactly.  In the ultra-relativistic limit their difference is suppressed by $m/E$, which is why weakly produced neutrinos are predominantly negative-helicity and antineutrinos predominantly positive-helicity.  The role of $U_1$ is therefore best described as a helicity-basis operation at fixed momentum, not as a statement about an invariant chiral transformation.

The operator $U_2$ is likewise a one-particle representation of charge conjugation at fixed helicity and up to conventional phases.  The standard field-theoretic charge-conjugation operation $\mathcal C$ maps creation operators for particles to creation operators for antiparticles and acts on spinors with the usual charge-conjugation matrix; it does not act on momentum or spin, so on one-particle states it preserves the helicity label,
\begin{equation}
\mathcal C\ket{\nu_h}=\ket{\bar\nu_h},
\qquad
\mathcal C\ket{\bar\nu_h}=\ket{\nu_h},
\label{eq:Caction}
\end{equation}
and therefore, in the basis \eqref{eq:basis} and with the standard phase convention,
\begin{equation}
\mathcal C=\tau_1\otimes\ident_2=U_2.
\label{eq:CisU2}
\end{equation}
Equation~\eqref{eq:CisU2} is used in Sec.~\ref{subsec:majorana} to express the Majorana condition as a projection.  It should not be confused with asserting that charge conjugation is a symmetry of the full interacting theory: the matrix $U_2$ is only the induced action on the restricted one-particle basis.  Whether that action is a symmetry depends on the interactions and on the conserved charges.  The Standard Model weak interactions are not invariant under charge conjugation, and exact lepton-number superselection would in any case forbid coherent rotations between $L=+1$ and $L=-1$ sectors.

The combined operator $U_3$ maps $\nu_-$ to $\bar\nu_+$ at fixed three-momentum.  It may have the same action on selected one-particle labels as part of a CPT-related transformation, but it is not the CPT operator: CPT is antiunitary and involves spacetime transformation properties, while $U_3$ is a unitary internal map in the fixed-momentum vector space.  This distinction is necessary for avoiding the overinterpretation that a basis transformation by itself establishes a physical lepton-number-violating process.

\section{Lepton-number superselection}
\label{sec:superselection}

The physical selection rule is encoded in the lepton-number operator \eqref{eq:L}.  Direct calculation gives
\begin{equation}
[U_1,L]=0,
\label{eq:U1L}
\end{equation}
whereas
\begin{equation}
\{U_2,L\}=0,
\qquad
\{U_3,L\}=0.
\label{eq:U23L}
\end{equation}
Equivalently, $U_1$ preserves the $L=\pm1$ sectors, while $U_2$ and $U_3$ exchange them:
\begin{equation}
L\ket{\nu_h}=+\ket{\nu_h},
\qquad
L\ket{\bar\nu_h}=-\ket{\bar\nu_h}.
\label{eq:Leigen}
\end{equation}
If lepton number is an exact conserved charge and is subject to superselection, physical observables must commute with $L$:
\begin{equation}
[O,L]=0.
\label{eq:superO}
\end{equation}
Under this condition, the charge-conjugating directions $U_2$ and $U_3$ are not observables within a single physical sector, even though they are well-defined as linear operators on the enlarged free space.

This is a different issue from ordinary flavor or mass coherence in neutrino oscillations.  A flavor neutrino produced in a weak process is a coherent superposition of different mass eigenstates because those components have the same total conserved charges and can interfere between production and detection \cite{Kayser1981,GiuntiKimLee1991,GiuntiKimLee1998}.  Lepton-number superselection, if exact, forbids a different kind of coherence: off-diagonal coherence between $L=+1$ and $L=-1$ sectors.  Thus ordinary oscillations among $\nu_i$ do not contradict the statement that $\nu$--$\bar\nu$ coherence requires $\Delta L\ne0$ dynamics.

\begin{figure}[t]
\centering
\includegraphics[width=0.98\columnwidth]{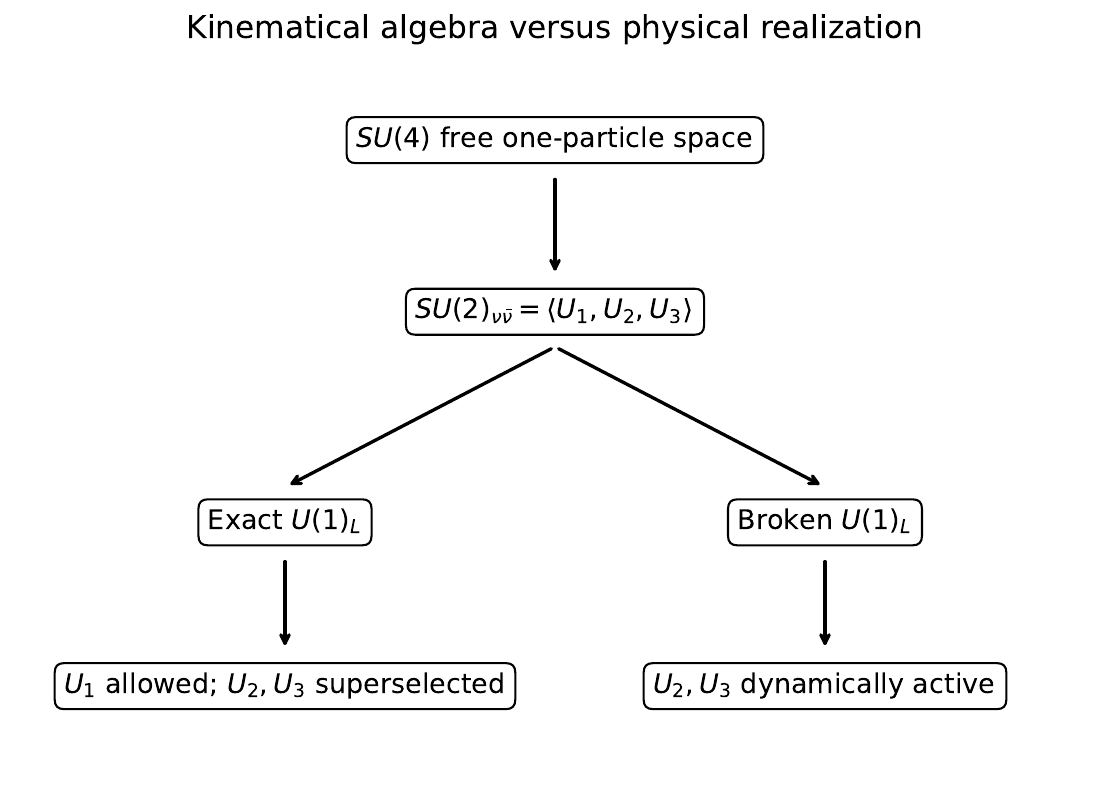}
\caption{Kinematical algebra versus physical realization.  The fixed-momentum free one-particle space admits an embedded $SU(2)_{\nu\bar\nu}$ generated by $U_1,U_2,U_3$.  If $U(1)_L$ is exact, $U_2$ and $U_3$ are superselected because they exchange lepton-number sectors.  If $U(1)_L$ is broken, the charge-conjugating directions become possible dynamical channels.}
\label{fig:flow}
\end{figure}

Figure~\ref{fig:flow} illustrates the separation between the algebraic embedding and the physical realization selected by lepton number.

The interpretation also clarifies the Dirac--Majorana distinction.  In a Dirac theory, $\nu$ and $\bar\nu$ are independent states distinguished by an exact charge.  In a Majorana theory, the labels conventionally associated with neutrino and antineutrino are not independent charge sectors but helicity and interaction labels of the same neutral particle.  The $U_3$ generator identifies the algebraic direction connecting the active pair $\nu_-$ and $\bar\nu_+$, but it does not by itself prove that the physical theory contains a Majorana mass.  A dynamical $\Delta L=2$ term is required.

\section{Density-matrix formulation}
\label{sec:density}

The tensor basis provides a compact density-matrix language for the four-state system.  A general fixed-momentum density matrix can be expanded as
\begin{equation}
\rho_{\mathbf p}=\frac{1}{4}\sum_{\mu,\nu=0}^3 r_{\mu\nu}\,\tau_\mu\otimes\sigma_\nu,
\label{eq:rhoexp}
\end{equation}
where
\begin{equation}
r_{\mu\nu}=\Tr\left[\rho_{\mathbf p}\,\tau_\mu\otimes\sigma_\nu\right].
\label{eq:rcoeff}
\end{equation}
A pure state $\ket{\psi}=a\ket{\nu_-}+b\ket{\bar\nu_+}$ corresponds to $\rho=\ket{\psi}\bra{\psi}$ and has off-diagonal $\nu$--$\bar\nu$ coherence proportional to $ab^*$.  An incoherent statistical mixture with probabilities $|a|^2$ and $|b|^2$ has the same diagonal populations but no such off-diagonal terms.  The algebra $SU(2)_{\nu\bar\nu}$ acts on state vectors and therefore on pure coherent superpositions; the density-matrix notation simply allows the same classification to include both pure states and mixed ensembles.

In block form,
\begin{equation}
\rho_{\mathbf p}=\begin{pmatrix}
\rho_{\nu\nu} & \rho_{\nu\bar\nu}\\
\rho_{\bar\nu\nu} & \rho_{\bar\nu\bar\nu}
\end{pmatrix}.
\label{eq:blockrho}
\end{equation}
The exact lepton-number superselection condition is
\begin{equation}
[\rho_{\mathbf p},L]=0.
\label{eq:rhosuper}
\end{equation}
In the expansion \eqref{eq:rhoexp}, this is equivalent to removing all $\tau_1$ and $\tau_2$ components:
\begin{equation}
r_{1\nu}=r_{2\nu}=0,
\qquad \nu=0,1,2,3.
\label{eq:coherencezero}
\end{equation}
Equivalently,
\begin{equation}
\rho_{\nu\bar\nu}=\rho_{\bar\nu\nu}=0.
\label{eq:offblocks}
\end{equation}
Thus the density-matrix formalism makes explicit which coherences are forbidden when lepton number is exact.

\begin{figure}[t]
\centering
\includegraphics[width=0.90\columnwidth]{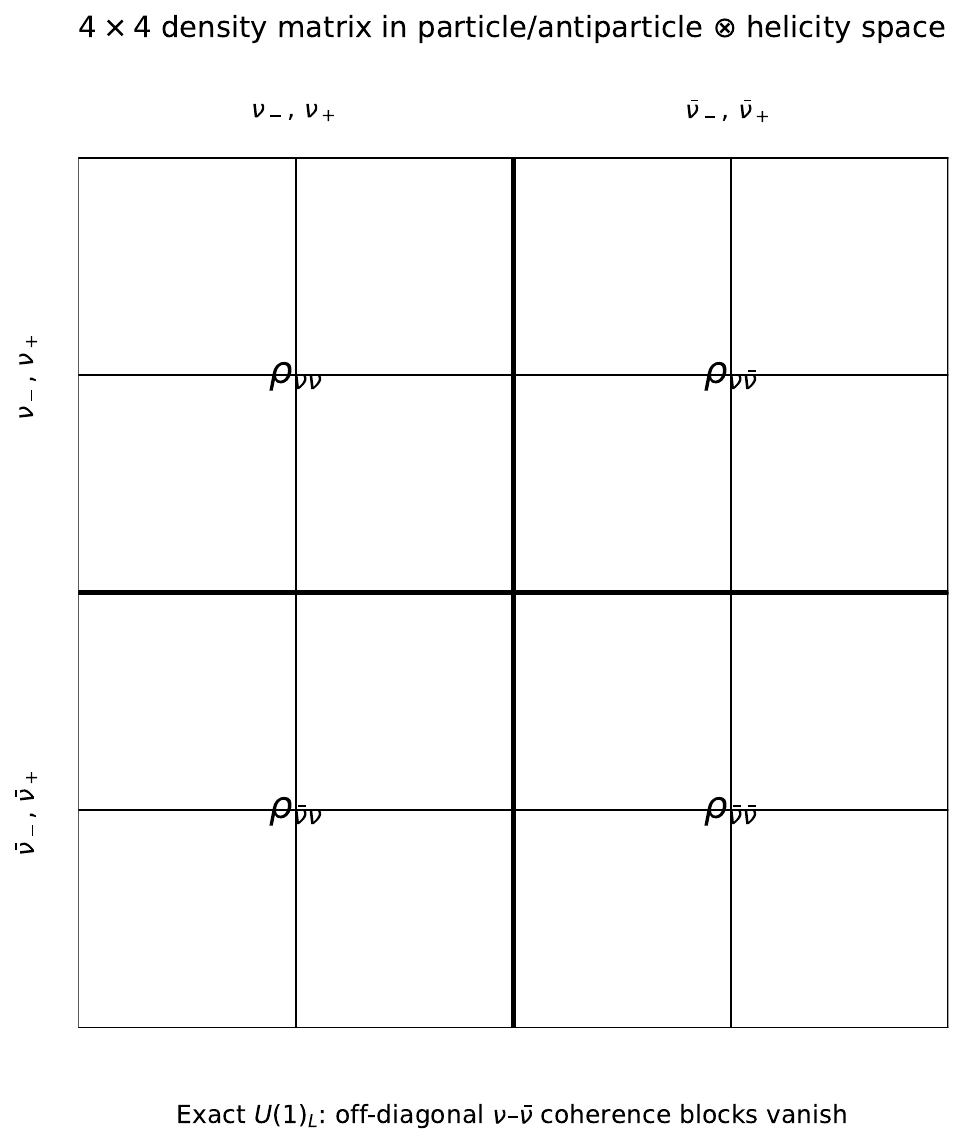}
\caption{Block structure of the $4\times4$ density matrix in particle--antiparticle $\otimes$ helicity space.  The off-diagonal blocks encode neutrino--antineutrino coherence.  Exact lepton-number superselection imposes $[\rho_{\mathbf p},L]=0$, which removes these blocks.}
\label{fig:blocks}
\end{figure}

The evolution of $\rho_{\mathbf p}$ in the collisionless limit is
\begin{equation}
i\frac{\dd \rho_{\mathbf p}}{\dd t}=[H_{\rm eff},\rho_{\mathbf p}],
\label{eq:vonNeumann}
\end{equation}
where a general effective Hamiltonian can be written as
\begin{equation}
H_{\rm eff}=E_{\mathbf p}\ident_4+
\sum_{\mu,\nu=0}^3 h_{\mu\nu}\,\tau_\mu\otimes\sigma_\nu.
\label{eq:Heff}
\end{equation}
The lepton-number condition is
\begin{equation}
[H_{\rm eff},L]=0.
\label{eq:HeffL}
\end{equation}
Because $L=\tau_3\otimes\ident_2$, Eq.~\eqref{eq:HeffL} eliminates all Hamiltonian components proportional to $\tau_1$ or $\tau_2$.  Therefore,
\begin{align}
H_{\Delta L=0}&=E_{\mathbf p}\ident_4+
\sum_{\nu=0}^3\left(h_{0\nu}\tau_0+h_{3\nu}\tau_3\right)\otimes\sigma_\nu,\label{eq:HL0}\\
H_{\Delta L\ne0}&=
\sum_{\nu=0}^3\left(h_{1\nu}\tau_1+h_{2\nu}\tau_2\right)\otimes\sigma_\nu.
\label{eq:HLV}
\end{align}
Equation~\eqref{eq:HLV} is the general one-particle form of a lepton-number-violating sector in this basis.  It does not replace a field-theoretic derivation of the microscopic interaction; it classifies the possible one-particle coherence channels once such a term is present.

\begin{table*}[t]
\caption{Effective Hamiltonian sectors in the $\tau\otimes\sigma$ basis, organized by the $\mathbb Z_2\times\mathbb Z_2$ grading of Eqs.~\eqref{eq:class00}--\eqref{eq:class21}.  $\Delta L$ is fixed by (anti)commutation with $L=\tau_3\otimes\ident_2$ and $\Delta h$ by (anti)commutation with $\mathcal H_h=\ident_2\otimes\sigma_3$.  The last column indicates whether the amplitude carries an explicit factor of the neutrino mass in the ultra-relativistic limit.  ``Active'' and ``sterile'' refer to the eigenvalue of $W=\tau_3\otimes\sigma_3$, Eq.~\eqref{eq:Wop}.}
\label{tab:classification}
\begin{ruledtabular}
\begin{tabular}{llllc}
Sector & $\Delta L$ & Helicity & Microscopic origin and channel & $\mathcal O(m_\nu/E)$?\\
\colrule
$\tau_0\otimes\sigma_0$ & $0$ & preserved & free propagation & ---\\
$\tau_3\otimes\sigma_0=L$ & $0$ & preserved & matter potential, opposite sign for $\nu$ and $\bar\nu$ & no\\
$\tau_{0}\otimes\sigma_3,\ \tau_3\otimes\sigma_3=W$ & $0$ & preserved & helicity-dependent energy shift (active versus sterile) & no\\
$\tau_{0,3}\otimes\sigma_{1,2}\ \ni U_1$ & $0$ & flipped & Dirac mass insertion; Dirac magnetic moment; active $\leftrightarrow$ sterile & yes\\
$\tau_{1,2}\otimes\sigma_{0,3}\ \ni U_2$ & $2$ & preserved & pseudo-Dirac mixing $\kappa_{\rm pD}=\Delta m^2/4E$; active $\leftrightarrow$ sterile & no\\
$\tau_{1,2}\otimes\sigma_{1,2}\ \ni U_3$ & $2$ & flipped & Majorana mass insertion; $0\nu\beta\beta$; spin coherence; active $\leftrightarrow$ active & yes\\
\end{tabular}
\end{ruledtabular}
\end{table*}

This classification is complementary to full neutrino quantum kinetic treatments, which include flavor, spin, collisions, and matter potentials in dense media.  Relativistic kinetic descriptions include neutrino and antineutrino degrees of freedom \cite{SiglRaffelt1993}; extended mean-field equations retain the $\nu$--$\bar\nu$ correlations that occupy the off-diagonal blocks of Eq.~\eqref{eq:blockrho} \cite{Volpe2013,SerreauVolpe2014,Kartavtsev2016}; and spin-coherence treatments have shown that spin terms can mediate neutrino--antineutrino transformation for Majorana neutrinos in anisotropic environments, or active--sterile transformation for Dirac neutrinos \cite{Vlasenko2014,Cirigliano2015}.  Supernova studies further indicate that the magnitude of such effects depends sensitively on the medium profile and dynamical conditions \cite{Tian2017}.  The present framework does not attempt to reproduce those kinetic equations.  Instead, it isolates the minimal internal algebra of a single massive neutral fermion at fixed momentum and classifies which sectors are compatible with exact or broken lepton number.

The restriction to a \emph{fixed} momentum should be kept in mind.  The coherences $\rho_{\nu\bar\nu}$ in Eq.~\eqref{eq:blockrho} are helicity coherences between $\nu$ and $\bar\nu$ of the same momentum.  They are not the pairing correlations between modes of opposite momenta, $\nu(\mathbf p)\,\bar\nu(-\mathbf p)$, which require an enlarged two-momentum space and which have been analyzed separately in the mean-field literature \cite{Volpe2013,SerreauVolpe2014,Kartavtsev2016}.  The classification below applies to the former and says nothing about the latter.

\section{Dirac, Majorana, and pseudo-Dirac limits}
\label{sec:limits}

The same algebraic space can describe different physical limits depending on the status of $U(1)_L$.

\subsection{Dirac limit}

In the strict Dirac limit, lepton number is exact.  The effective Hamiltonian obeys
\begin{equation}
[H_{\rm eff},L]=0,
\label{eq:DiracCondition}
\end{equation}
and an initially block-diagonal density matrix remains block diagonal:
\begin{equation}
\rho_{\nu\bar\nu}(0)=0
\quad\Rightarrow\quad
\rho_{\nu\bar\nu}(t)=0.
\label{eq:DiracBlock}
\end{equation}
The internal algebra still exists at the level of the enlarged free vector space, but the $U_2$ and $U_3$ directions do not correspond to physical coherent rotations.  They are forbidden by the same logic that forbids coherent rotations between exactly distinct charge sectors.

If sterile right-handed neutrinos are present, the field content required for a Dirac mass is enlarged, but the statement above applies mass eigenstate by mass eigenstate once the Dirac pair has been formed.  Additional sterile flavors enlarge the flavor Hilbert space and therefore the full kinetic problem, but they do not remove the lepton-number distinction between Dirac particles and antiparticles unless $\Delta L=2$ terms are also introduced.

\subsection{Majorana limit}
\label{subsec:majorana}

In a Majorana theory, there is no independent conserved lepton-number charge distinguishing neutrino and antineutrino states.  The observed distinction between neutrino and antineutrino is then tied to helicity and weak-interaction selection rather than to an exact particle--antiparticle charge.  In the present language, the $\tau_1$ and $\tau_2$ directions are no longer excluded by a fundamental $U(1)_L$ superselection rule.

The identification \eqref{eq:CisU2} makes the Majorana condition explicit.  A Majorana particle is its own antiparticle, $\mathcal C\ket{\psi}=\ket{\psi}$ up to a Majorana phase, so the physical Majorana one-particle space is the $U_2=+1$ eigenspace of $\mathcal H_{\mathbf p}$,
\begin{equation}
\mathcal H_{\mathbf p}^{\rm Maj}=\Pi_+\,\mathcal H_{\mathbf p},
\qquad
\Pi_\pm=\tfrac12\left(\ident_4\pm U_2\right),
\label{eq:majproj}
\end{equation}
spanned by the two states
\begin{equation}
\ket{\chi_h}=\frac{1}{\sqrt2}\left(\ket{\nu_h}+\ket{\bar\nu_h}\right),
\qquad h=\pm.
\label{eq:majstates}
\end{equation}
The four-state space collapses to two states labeled only by helicity, as it must for a single neutral fermion.  The off-diagonal block $\rho_{\nu\bar\nu}$, which in the Dirac language is a lepton-number-violating coherence between two distinct particles, becomes on $\mathcal H_{\mathbf p}^{\rm Maj}$ nothing more than the ordinary helicity density matrix of one particle.  Two different names, and two different intuitions, attach to the same matrix element.

The projection also determines what remains of the pseudospin algebra.  Because $\{U_1,U_2\}=\{U_3,U_2\}=0$, both $U_1$ and $U_3$ map $\Pi_+\mathcal H_{\mathbf p}$ onto $\Pi_-\mathcal H_{\mathbf p}$:
\begin{equation}
\Pi_+\,U_1\,\Pi_+=\Pi_+\,U_3\,\Pi_+=0.
\label{eq:U13out}
\end{equation}
Meanwhile $[\tau_0\otimes\sigma_i,U_2]=0$, so the Wigner spin generators \eqref{eq:WignerJ} preserve $\Pi_\pm$.  In the Majorana theory, therefore, $U_2$ is not a transformation between distinct states but the constraint defining the physical subspace; $U_1$ and $U_3$ leave that subspace; and the only $SU(2)$ acting within it is the ordinary spin little group.  The pseudospin algebra is absent from the physical sector in both limits, though for different reasons: superselected in the Dirac case, projected out in the Majorana case.

None of this is a proof of Majorana dynamics.  A Majorana mass term or an effective $\Delta L=2$ interaction supplies the Hamiltonian component that activates the corresponding coherence.  Without such a term, the kinematical generator remains only a possible direction in the enlarged algebra.

\subsection{Pseudo-Dirac limit and mass splitting}
\label{subsec:pseudodirac}

A pseudo-Dirac neutrino can be viewed as a pair of nearly degenerate Majorana states whose splitting is controlled by small lepton-number-violating mass terms \cite{KobayashiLim2001,Beacom2004}.  It is convenient to use two left-handed Weyl fields,
\begin{equation}
\psi_1=\nu_L\ \ (L=+1),
\qquad
\psi_2=N_R^{\,c}\ \ (L=-1),
\label{eq:weylfields}
\end{equation}
with the mass term $-\tfrac12\psi^{T}\mathcal M\psi+{\rm h.c.}$ and
\begin{equation}
\mathcal M=\begin{pmatrix}
m_L & m_D\\
m_D & m_R
\end{pmatrix},
\label{eq:massmatrix}
\end{equation}
where $m_D$ is lepton-number conserving and $m_L,m_R$ violate lepton number by two units.  In the pseudo-Dirac regime $|m_L|,|m_R|\ll m_D$.

A left-handed Weyl field annihilates a negative-helicity particle and creates a positive-helicity antiparticle.  Hence
\begin{align}
\psi_1:&\quad\text{annihilates }\ket{\nu_-},\ \ \text{creates }\ket{\bar\nu_+},\label{eq:psi1states}\\
\psi_2:&\quad\text{annihilates }\ket{\bar\nu_-},\ \ \text{creates }\ket{\nu_+}.\label{eq:psi2states}
\end{align}
The weakly active states are thus $\{\ket{\nu_-},\ket{\bar\nu_+}\}$ and the sterile states $\{\ket{\nu_+},\ket{\bar\nu_-}\}$, in agreement with the eigenvalues of $W$ in Eq.~\eqref{eq:Wop}.

Two distinct $\Delta L=2$ structures follow, and they lie in \emph{different} grading classes.

\emph{(i) The Lagrangian insertion.}  The terms $m_L\psi_1\psi_1$ and $m_R\psi_2\psi_2$ connect the states annihilated and created by the same Weyl field, that is $\ket{\nu_-}\to\ket{\bar\nu_+}$ and $\ket{\bar\nu_-}\to\ket{\nu_+}$.  For the CP-conserving case $m_L=m_R=\mu_M$ this is
\begin{equation}
H_{\Delta L=2}^{\rm ins}=\mu_M\,\tau_1\otimes\sigma_1,
\label{eq:pDinsertion}
\end{equation}
a helicity-flipping term in the $U_3$ class \eqref{eq:class21}, with coefficient linear in the small mass $\mu_M$.

\emph{(ii) The induced one-particle mixing.}  Propagation is governed instead by the second-order combination $\mathcal M\mathcal M^\dagger$.  In the ultra-relativistic limit the one-particle Hamiltonian in the negative-helicity particle sector $(\ket{\nu_-},\ket{\bar\nu_-})$ is $E_{\mathbf p}\ident_2+\mathcal M\mathcal M^\dagger/2E_{\mathbf p}$, and in the positive-helicity sector $(\ket{\bar\nu_+},\ket{\nu_+})$ it is $E_{\mathbf p}\ident_2+\mathcal M^{T}\mathcal M^{*}/2E_{\mathbf p}$.  With $m_L=m_R=\mu_M$ real,
\begin{equation}
\mathcal M\mathcal M^\dagger=
\begin{pmatrix}
m_D^2+\mu_M^2 & 2m_D\mu_M\\
2m_D\mu_M & m_D^2+\mu_M^2
\end{pmatrix},
\label{eq:MMdag}
\end{equation}
and assembling both helicity sectors in the basis \eqref{eq:basis} gives
\begin{equation}
H_{\rm pD}=E_{\mathbf p}\ident_4+\kappa_{\rm pD}\,\tau_1\otimes\ident_2
=E_{\mathbf p}\ident_4+\kappa_{\rm pD}\,U_2,
\label{eq:pDsimple}
\end{equation}
with
\begin{equation}
\kappa_{\rm pD}\simeq \frac{m_D\mu_M}{E_{\mathbf p}}=\frac{\Delta m^2}{4E_{\mathbf p}},
\qquad
\Delta m^2\equiv m_+^2-m_-^2\simeq 4m_D\mu_M,
\label{eq:kappapd}
\end{equation}
where $m_\pm\simeq m_D\pm\mu_M$.  Explicitly,
\begin{equation}
H_{\rm pD}=\begin{pmatrix}
E_{\mathbf p}&0&\kappa_{\rm pD}&0\\
0&E_{\mathbf p}&0&\kappa_{\rm pD}\\
\kappa_{\rm pD}&0&E_{\mathbf p}&0\\
0&\kappa_{\rm pD}&0&E_{\mathbf p}
\end{pmatrix}.
\label{eq:HpDmatrix}
\end{equation}
The two doublets have one-particle energies $E_{\mathbf p}\pm\kappa_{\rm pD}$, so that $E_+-E_-=2\kappa_{\rm pD}=\Delta m^2/2E_{\mathbf p}$, as required.

Equations~\eqref{eq:pDinsertion} and \eqref{eq:pDsimple} are the central point of this subsection.  The Lagrangian mass insertion is helicity-flipping and belongs to the $U_3$ class; the induced one-particle energy mixing is helicity-\emph{preserving} and belongs to the $U_2$ class.  This is the one-particle counterpart of the familiar statement that a mass insertion flips chirality whereas propagation over a macroscopic baseline does not.  It also settles which channel carries the pseudo-Dirac observable: the physical transition is $\ket{\nu_-}\to\ket{\bar\nu_-}$, an active-to-sterile conversion at fixed helicity generated by $U_2$, and \emph{not} the active-to-active channel $\ket{\nu_-}\to\ket{\bar\nu_+}$ generated by $U_3$.  Attaching the coefficient $\Delta m^2/4E$ to the $U_3$ direction would misidentify both the final state and its mass dependence.

A more general lepton-number-violating propagation term in the same grading class can be parameterized as
\begin{equation}
H_{\rm pD}^{\rm gen}=
\left(\cos\phi\,\tau_1+\sin\phi\,\tau_2\right)
\otimes
\left(\kappa_{\rm pD}\,\sigma_0+\kappa_3\,\sigma_3\right),
\label{eq:pDgeneral}
\end{equation}
where $\phi$ encodes a CP-violating phase in the lepton-number-violating sector and $\kappa_3$ allows the two helicities to be split.  The essential point is independent of this parameterization: pseudo-Dirac propagation corresponds to small but nonzero $\tau_{1,2}\otimes\sigma_{0,3}$ components of the one-particle Hamiltonian.

\subsection{Helicity selection within the $\Delta L=2$ sector}
\label{sec:helicitySelection}

The grading of Sec.~\ref{subsec:grading} splits the lepton-number-violating sector into two classes, and the split has direct phenomenological content.

\emph{Helicity-preserving, $\tau_{1,2}\otimes\sigma_{0,3}$ (the $U_2$ class).}  By Eq.~\eqref{eq:UW} these terms connect an active state to a sterile one at fixed helicity.  No chirality flip is required, so the amplitude carries no explicit factor of $m_\nu/E$.  The pseudo-Dirac coefficient $\kappa_{\rm pD}$ produces maximal vacuum mixing between $\ket{\nu_-}$ and $\ket{\bar\nu_-}$, and the observable is controlled by the accumulated phase $\kappa_{\rm pD}L$ rather than by a suppressed amplitude.  This is why pseudo-Dirac splittings far below any laboratory sensitivity remain accessible over astrophysical baselines \cite{Beacom2004}.

\emph{Helicity-flipping, $\tau_{1,2}\otimes\sigma_{1,2}$ (the $U_3$ class).}  These terms connect two active states, $\nu_-\leftrightarrow\bar\nu_+$, or two sterile ones.  A chirality flip is required, so in the ultra-relativistic limit the amplitude carries an explicit factor of the neutrino mass, as in Eq.~\eqref{eq:pDinsertion}.  This is the algebraic origin of the $m_\nu$ dependence of the neutrinoless double-beta decay amplitude, summarized by $m_{\beta\beta}$ \cite{SchechterValle1982,BilenkyGiunti2015}, and of the $\mathcal O(m/E)$ suppression of the spin-coherence terms in neutrino quantum kinetics \cite{Vlasenko2014,Cirigliano2015,Tian2017}.

The grading also yields, as an immediate corollary, the known dichotomy between Dirac and Majorana spin coherence.  A helicity-flipping one-particle term must lie in one of exactly two classes.  If it is lepton-number conserving, $\tau_{0,3}\otimes\sigma_{1,2}$, it converts an active neutrino into a sterile one; this is the Dirac case, and $U_1$ is its representative.  If it violates lepton number, $\tau_{1,2}\otimes\sigma_{1,2}$, it converts a neutrino into an antineutrino; this is the Majorana case, and $U_3$ is its representative.  There is no third possibility.  The statement that medium-induced spin coherence produces $\nu\to\bar\nu$ transformation for Majorana neutrinos and active--sterile transformation for Dirac neutrinos \cite{Vlasenko2014,Cirigliano2015} is, in this language, a one-line consequence of Eqs.~\eqref{eq:class01} and \eqref{eq:class21}.

\subsection{Massless limit}
\label{subsec:massless}

The strictly massless limit is singular for the present discussion in two respects.  First, the Wigner little group changes from the massive $SU(2)$ spin little group to the massless $ISO(2)$ little group, with physical one-particle states labeled by helicity rather than by a rest-frame spin multiplet \cite{Wigner1939,WeinbergQFT1}.  Second, in the absence of mass terms the left- and right-chiral fields decouple.  The algebraic four-state construction can still be used as a formal basis if both helicities and both charge-conjugate states are included, but the interpretation as a massive fixed-momentum degeneracy algebra no longer applies in the same way.  The physically relevant neutrino limit is therefore not $m=0$, but $m/E\ll1$, where wrong-helicity components and Dirac--Majorana differences are suppressed but not conceptually absent.

\section{Minimal two-state models for the $\Delta L=2$ channels}
\label{sec:minimal}

Each of the two lepton-number-violating classes reduces, at fixed momentum, to a two-state problem.  For the helicity-preserving ($U_2$) channel the relevant subspace is
\begin{equation}
\mathcal H_{U_2}=\mathrm{span}\left\{\ket{\nu_-},\ket{\bar\nu_-}\right\},
\label{eq:U2subspace}
\end{equation}
whereas for the helicity-flipping ($U_3$) channel it is
\begin{equation}
\mathcal H_{U_3}=\mathrm{span}\left\{\ket{\nu_-},\ket{\bar\nu_+}\right\}.
\label{eq:activeSubspace}
\end{equation}
In either case the most general Hermitian two-state Hamiltonian after subtracting an irrelevant trace is
\begin{equation}
H_{\nu\bar\nu}=\begin{pmatrix}
\delta/2 & \kappa\\
\kappa^* & -\delta/2
\end{pmatrix}.
\label{eq:twostateH}
\end{equation}
Here $\delta$ is an effective splitting between the two states and $\kappa$ is the lepton-number-violating mixing amplitude.  Writing $\ket{1}$ and $\ket{2}$ for the two basis states, the transition probability is
\begin{equation}
P_{1\rightarrow2}(t)=
\frac{|\kappa|^2}{|\kappa|^2+\delta^2/4}
\sin^2\left(\sqrt{|\kappa|^2+\delta^2/4}\,t\right).
\label{eq:prob}
\end{equation}
Equation~\eqref{eq:prob} is deliberately minimal, but it captures the main physical point.  If $\kappa=0$, the algebraic direction exists but the transition probability vanishes:
\begin{equation}
\kappa=0\quad\Rightarrow\quad P_{1\rightarrow2}(t)=0.
\label{eq:kappazero}
\end{equation}
If $\kappa\ne0$, the same direction becomes dynamically active.  A nonzero splitting $\delta$ suppresses coherent conversion unless $|\kappa|$ is comparable to or larger than $|\delta|/2$.  The two channels differ in what $\kappa$ and $\delta$ are.

\begin{figure}[t]
\centering
\includegraphics[width=0.98\columnwidth]{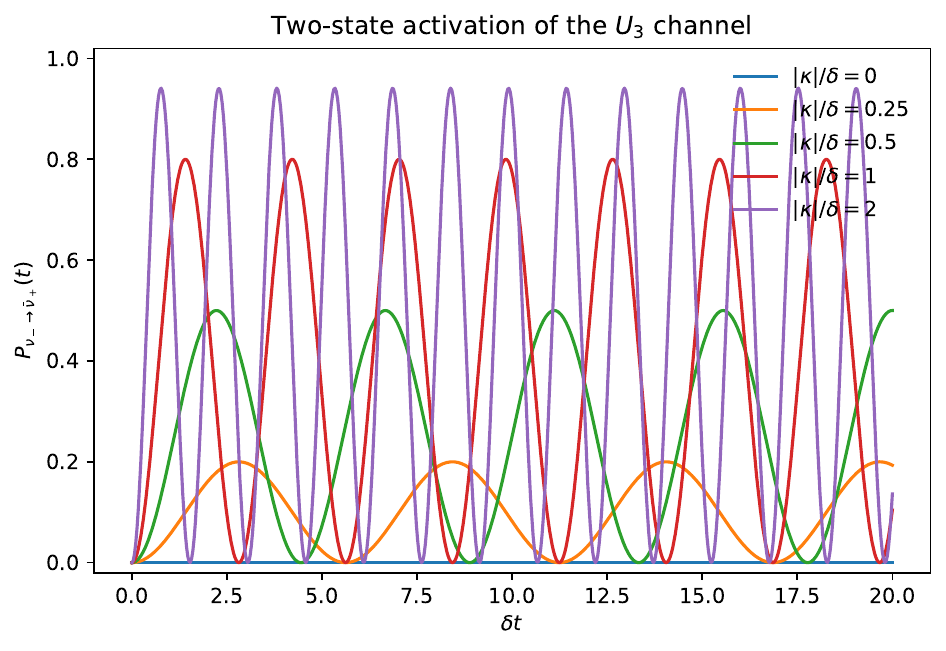}
\caption{Transition probability in the minimal two-state model, Eq.~\eqref{eq:prob}, applicable to either $\Delta L=2$ channel with the appropriate $(\kappa,\delta)$.  The algebraic direction exists even for $\kappa=0$, but no transition occurs unless a lepton-number-violating mixing amplitude is present.  For the pseudo-Dirac ($U_2$) channel in vacuum $\delta=0$ and the mixing is maximal; for the $U_3$ channel in a medium $\delta\simeq2V$ and conversion is strongly suppressed.}
\label{fig:transition}
\end{figure}

\subsection{The $U_2$ channel: pseudo-Dirac oscillation}

Here $\kappa=\kappa_{\rm pD}=\Delta m^2/4E$ from Eq.~\eqref{eq:kappapd}.  In vacuum $\delta=0$, because $\ket{\nu_-}$ and $\ket{\bar\nu_-}$ are degenerate at fixed momentum, so the mixing is maximal and
\begin{equation}
P_{\nu_-\rightarrow\bar\nu_-}(L)
=\sin^2\left(\frac{\Delta m^2 L}{4E}\right).
\label{eq:pDprob}
\end{equation}
This is the active-to-sterile conversion that makes astrophysical baselines sensitive to extremely small pseudo-Dirac splittings \cite{KobayashiLim2001,Beacom2004}.  In matter, $\ket{\nu_-}$ is weakly interacting and $\ket{\bar\nu_-}$ is not, so $\delta$ is the matter potential of the active state and conversion is suppressed once $|\delta|\gtrsim2|\kappa_{\rm pD}|$.

\subsection{The $U_3$ channel: neutrino--antineutrino coherence}

Here both states are weakly active, and $\kappa$ is a helicity-flipping $\Delta L=2$ amplitude carrying an explicit factor of the neutrino mass.  The two states are again degenerate in vacuum, but in a medium $\ket{\nu_-}$ and $\ket{\bar\nu_+}$ acquire matter potentials of opposite sign, so $\delta\simeq2V$ and the conversion probability is bounded by $|\kappa|^2/(|\kappa|^2+V^2)$.  Since $|\kappa|=\mathcal O(m_\nu)$, the suppression is severe.  This is the one-particle image of the strong suppression of helicity coherence found in dense-medium calculations \cite{Tian2017} and of the absence of mass-independent collective $\nu$--$\bar\nu$ conversion \cite{FiorilloRaffeltSigl2024}.

In neutrinoless double-beta decay the relevant $\Delta L=2$ object is not a universal propagation coefficient at all, but the process-dependent Majorana insertion in the virtual-neutrino amplitude, conventionally summarized by the effective mass $m_{\beta\beta}=|\sum_i U_{ei}^2m_i|$ for light-neutrino exchange \cite{SchechterValle1982,BilenkyGiunti2015}.  Both channels activate $\tau_{1,2}$ directions, but the mapping from the microscopic theory to $\kappa$ is observable dependent.

The same caution applies to heavy-neutrino decays and nonrelativistic probes.  Nonrelativistic neutrinos can display Dirac--Majorana differences that are invisible in the ultra-relativistic confusion limit, for example through angular distributions in heavy-neutrino decays \cite{Balantekin2019}.  The present two-level models are not a replacement for those process calculations.  Their role is to make clear which one-particle sector a lepton-number-violating amplitude must populate.

\section{Discussion}
\label{sec:discussion}

The main result can be summarized in five statements.

First, the four-state massive neutral-fermion space at fixed momentum admits an internal $SU(2)$ algebra generated by $U_1,U_2,U_3$.  This algebra preserves the free four-momentum and is therefore little-group-like in a limited sense.  However, because $U_2$ and $U_3$ exchange particle and antiparticle sectors, the algebra is not the ordinary Wigner spin little group.  It is an internal pseudospin algebra embedded in the larger degeneracy algebra of the free one-particle space.

Second, the triplet is distinguished by a grading rather than by uniqueness.  The full $SU(4)$ space contains many $SU(2)$ subalgebras, but lepton number $L$ and the helicity label $\mathcal H_h$ are commuting involutions that sort the sixteen tensor operators into four classes, and $U_1,U_2,U_3$ are one representative of each nontrivial class.  The identity $U_1=2LJ_2$ makes precise the sense in which even the helicity generator is not a spin rotation: it is the lepton-number-weighted one.

Third, lepton number selects the physical content.  The helicity-flip generator $U_1$ commutes with $L$, whereas the charge-conjugating generators $U_2$ and $U_3$ anticommute with it.  Therefore, if lepton number is exact and superselected, $U_2$ and $U_3$ cannot be physical coherent rotations within a single sector.  This is the precise sense in which the existence of the algebra does not by itself favor Majorana neutrinos.  It identifies possible directions in state space, not a dynamical law.

Fourth, when lepton number is broken, helicity decides what is observed.  Hamiltonian components proportional to $\tau_1$ or $\tau_2$ generate neutrino--antineutrino coherence, but they fall into two grading classes with different physics.  The helicity-preserving class $\tau_{1,2}\otimes\sigma_{0,3}$, containing $U_2$, connects active to sterile states, carries the pseudo-Dirac coefficient $\kappa_{\rm pD}=\Delta m^2/4E$, and is not amplitude-suppressed.  The helicity-flipping class $\tau_{1,2}\otimes\sigma_{1,2}$, containing $U_3$, connects active to active states and carries an explicit factor of the neutrino mass.  Attaching the pseudo-Dirac coefficient to the $U_3$ direction misidentifies both the final state and its mass dependence.

Fifth, in the Majorana theory the pseudospin algebra is not merely superselected but absent.  Charge conjugation on one-particle states is $U_2$, the Majorana condition is the projection onto its $+1$ eigenspace, and $U_1$ and $U_3$ map that subspace onto its complement.  What remains acting within the physical two-state space is the ordinary Wigner spin algebra.  The off-diagonal block $\rho_{\nu\bar\nu}$, a lepton-number-violating coherence in Dirac language, is on the Majorana subspace nothing but the helicity density matrix of a single neutral particle.  Part of the persistent difficulty of the Dirac--Majorana comparison is that the same matrix element carries two names.

This perspective complements amplitude-level studies of Dirac--Majorana distinguishability \cite{LiWilczek1982,KayserShrock1982,Rodejohann2017,Kim2023}.  Those works ask when measurable processes can distinguish Dirac and Majorana neutrinos.  Here the question is more structural: which internal one-particle directions are compatible with exact lepton number, and which require $\Delta L=2$ dynamics?  The answer is encoded in Eqs.~\eqref{eq:U1L} and \eqref{eq:U23L}.

The result also clarifies the interpretation of the earlier little-group construction \cite{Romero2021}.  The transformation connecting $\nu_-$ and $\bar\nu_+$ is physically significant, but not because its existence proves the Majorana nature of neutrinos.  Its significance is that it identifies the algebraic channel that a Majorana mass term, pseudo-Dirac splitting, or other lepton-number-violating interaction would activate.  This conservative interpretation is consistent with the broader lesson that algebraic spinor properties and unitary basis relations must be distinguished from new physical particle content \cite{Romero2023}.

\section{Conclusions}
\label{sec:conclusion}

We have reformulated helicity and charge-conjugation transformations of massive neutral-fermion one-particle states as an internal pseudospin algebra.  In the fixed-momentum basis $(\nu_-,\nu_+,\bar\nu_-,\bar\nu_+)$, the three operators
\begin{equation}
U_1=\tau_3\otimes\sigma_2,
\qquad
U_2=\tau_1\otimes\ident_2,
\qquad
U_3=\tau_2\otimes\sigma_2
\end{equation}
are Hermitian, unitary, and close an $SU(2)$ algebra.  The lepton-number operator $L=\tau_3\otimes\ident_2$ separates their physical roles: $U_1$ preserves lepton number, while $U_2$ and $U_3$ exchange lepton-number sectors.

The resulting interpretation is that the $SU(2)$ algebra is kinematically present in the enlarged free one-particle space, but only its lepton-number-preserving part is physically available under exact $U(1)_L$ superselection.  The charge-conjugating directions become dynamically meaningful only when lepton number is broken.  The density-matrix formulation makes this statement explicit: exact lepton number removes all $\tau_1$ and $\tau_2$ coherences, while $\Delta L=2$ Hamiltonian terms restore them.

The pseudo-Dirac mapping shows how this abstract classification connects to a standard mass-splitting observable, and where the connection is easy to misplace.  A small Majorana mass insertion is a helicity-flipping term of order $\mu_M$ in the $U_3$ class, but the one-particle \emph{propagation} coefficient it induces, $\kappa_{\rm pD}\simeq\Delta m^2/(4E)$, is helicity-preserving and lies in the $U_2$ class.  The pseudo-Dirac observable is therefore the active-to-sterile transition $\nu_-\to\bar\nu_-$ with the usual long-baseline phase $\Delta m^2L/(4E)$, not an active-to-active $\nu\to\bar\nu$ conversion.  Other observables, such as neutrinoless double-beta decay or heavy-neutrino decays, activate the lepton-number-violating sectors through different process-dependent amplitudes.

More generally, the $\mathbb Z_2\times\mathbb Z_2$ grading by lepton number and helicity, rather than lepton number alone, is what determines the physical content of a one-particle operator.  It fixes which $\Delta L=2$ amplitudes carry an explicit factor of the neutrino mass, and it reproduces in one line the dichotomy between Dirac active--sterile and Majorana neutrino--antineutrino spin coherence.

Thus the transformation $U_3$ should not be read as evidence by itself that neutrinos are Majorana particles.  Rather, it is the internal pseudospin direction associated with the physically important channel $\nu_-\leftrightarrow\bar\nu_+$.  Its activation is a dynamical question, controlled by Majorana masses, pseudo-Dirac splittings, medium-induced spin coherence, or other lepton-number-violating interactions.

\begin{acknowledgments}
The author thanks the Departamento de Ciencias Naturales, Universidad Aut\'onoma Metropolitana Unidad Cuajimalpa, for institutional support.
\end{acknowledgments}

\appendix

\section{Matrix checks}
\label{app:checks}

For completeness, we give the elementary algebraic checks.  From Eqs.~\eqref{eq:U1}--\eqref{eq:U3},
\begin{equation}
U_1U_2=(\tau_3\tau_1)\otimes\sigma_2=i\tau_2\otimes\sigma_2=iU_3,
\end{equation}
whereas
\begin{equation}
U_2U_1=(\tau_1\tau_3)\otimes\sigma_2=-i\tau_2\otimes\sigma_2=-iU_3.
\end{equation}
Therefore
\begin{equation}
[U_1,U_2]=2iU_3.
\end{equation}
Cyclic permutations give the remaining commutators.  The lepton-number relations follow similarly:
\begin{align}
[U_1,L]&=[\tau_3\otimes\sigma_2,\tau_3\otimes\ident_2]=0,\\
\{U_2,L\}&=\{\tau_1,\tau_3\}\otimes\ident_2=0,\\
\{U_3,L\}&=\{\tau_2,\tau_3\}\otimes\sigma_2=0.
\end{align}

\section{Transition probability}
\label{app:transition}

The Hamiltonian \eqref{eq:twostateH} has eigenfrequency
\begin{equation}
\Omega=\sqrt{|\kappa|^2+\delta^2/4}.
\end{equation}
The evolution operator is
\begin{equation}
U(t)=\cos(\Omega t)\ident_2-i\frac{\sin(\Omega t)}{\Omega}H_{\nu\bar\nu}.
\end{equation}
Starting from $\ket{1}$, the transition amplitude to $\ket{2}$ is
\begin{equation}
\mathcal A_{1\rightarrow2}(t)=-i\frac{\kappa^*}{\Omega}\sin(\Omega t),
\end{equation}
which gives Eq.~\eqref{eq:prob}.  The same expression applies to the $U_2$ channel with $\ket{1}=\ket{\nu_-}$, $\ket{2}=\ket{\bar\nu_-}$ and to the $U_3$ channel with $\ket{2}=\ket{\bar\nu_+}$.


\begin{thebibliography}{99}

\bibitem{Majorana1937}
E. Majorana, ``Teoria simmetrica dell'elettrone e del positrone,'' Nuovo Cimento \textbf{14}, 171--184 (1937), \href{https://doi.org/10.1007/BF02961314}{doi:10.1007/BF02961314}.

\bibitem{GiuntiKim2007}
C. Giunti and C. W. Kim, \textit{Fundamentals of Neutrino Physics and Astrophysics} (Oxford University Press, Oxford, 2007).

\bibitem{MohapatraPal2004}
R. N. Mohapatra and P. B. Pal, \textit{Massive Neutrinos in Physics and Astrophysics}, 3rd ed. (World Scientific, Singapore, 2004), \href{https://doi.org/10.1142/5024}{doi:10.1142/5024}.

\bibitem{SchechterValle1982}
J. Schechter and J. W. F. Valle, ``Neutrinoless Double-$\beta$ Decay in $SU(2)\times U(1)$ Theories,'' Phys. Rev. D \textbf{25}, 2951--2954 (1982), \href{https://doi.org/10.1103/PhysRevD.25.2951}{doi:10.1103/PhysRevD.25.2951}.

\bibitem{BilenkyGiunti2015}
S. M. Bilenky and C. Giunti, ``Neutrinoless Double-Beta Decay: A Probe of Physics Beyond the Standard Model,'' Int. J. Mod. Phys. A \textbf{30}, 1530001 (2015), \href{https://doi.org/10.1142/S0217751X1530001X}{doi:10.1142/S0217751X1530001X}.

\bibitem{Hirsch2018}
M. Hirsch, R. Srivastava, and J. W. F. Valle, ``Can One Ever Prove That Neutrinos Are Dirac Particles?,'' Phys. Lett. B \textbf{781}, 302--305 (2018), \href{https://doi.org/10.1016/j.physletb.2018.03.073}{doi:10.1016/j.physletb.2018.03.073}.

\bibitem{LiWilczek1982}
L.-F. Li and F. Wilczek, ``Physical Processes Involving Majorana Neutrinos,'' Phys. Rev. D \textbf{25}, 143--146 (1982), \href{https://doi.org/10.1103/PhysRevD.25.143}{doi:10.1103/PhysRevD.25.143}.

\bibitem{KayserShrock1982}
B. Kayser and R. E. Shrock, ``Distinguishing Between Dirac and Majorana Neutrinos in Neutral-Current Reactions,'' Phys. Lett. B \textbf{112}, 137--142 (1982), \href{https://doi.org/10.1016/0370-2693(82)90314-8}{doi:10.1016/0370-2693(82)90314-8}.

\bibitem{Rodejohann2017}
W. Rodejohann, X.-J. Xu, and C. E. Yaguna, ``Distinguishing Between Dirac and Majorana Neutrinos in the Presence of General Interactions,'' J. High Energy Phys. \textbf{05}, 024 (2017), \href{https://doi.org/10.1007/JHEP05(2017)024}{doi:10.1007/JHEP05(2017)024}, arXiv:1702.05721 [hep-ph].

\bibitem{Kim2023}
C. S. Kim, ``Practical Dirac Majorana Confusion Theorem: Issues and Applicability,'' Eur. Phys. J. C \textbf{83}, 972 (2023), \href{https://doi.org/10.1140/epjc/s10052-023-12156-9}{doi:10.1140/epjc/s10052-023-12156-9}, arXiv:2307.05654 [hep-ph].

\bibitem{Kayser1981}
B. Kayser, ``On the Quantum Mechanics of Neutrino Oscillation,'' Phys. Rev. D \textbf{24}, 110--116 (1981), \href{https://doi.org/10.1103/PhysRevD.24.110}{doi:10.1103/PhysRevD.24.110}.

\bibitem{GiuntiKimLee1991}
C. Giunti, C. W. Kim, and U. W. Lee, ``When Do Neutrinos Really Oscillate? Quantum Mechanics of Neutrino Oscillations,'' Phys. Rev. D \textbf{44}, 3635--3640 (1991), \href{https://doi.org/10.1103/PhysRevD.44.3635}{doi:10.1103/PhysRevD.44.3635}.

\bibitem{GiuntiKimLee1998}
C. Giunti, C. W. Kim, and U. W. Lee, ``When Do Neutrinos Cease to Oscillate?,'' Phys. Lett. B \textbf{421}, 237--244 (1998), \href{https://doi.org/10.1016/S0370-2693(98)00014-8}{doi:10.1016/S0370-2693(98)00014-8}.

\bibitem{Wigner1939}
E. P. Wigner, ``On Unitary Representations of the Inhomogeneous Lorentz Group,'' Ann. Math. \textbf{40}, 149--204 (1939), \href{https://doi.org/10.2307/1968551}{doi:10.2307/1968551}.

\bibitem{Weinberg1964}
S. Weinberg, ``Feynman Rules for Any Spin,'' Phys. Rev. \textbf{134}, B882--B896 (1964), \href{https://doi.org/10.1103/PhysRev.134.B882}{doi:10.1103/PhysRev.134.B882}.

\bibitem{WeinbergQFT1}
S. Weinberg, \textit{The Quantum Theory of Fields. Vol. 1: Foundations} (Cambridge University Press, Cambridge, 1995), \href{https://doi.org/10.1017/CBO9781139644167}{doi:10.1017/CBO9781139644167}.

\bibitem{Romero2021}
R. Romero, ``Little Group Generators for Dirac Neutrino One-Particle States,'' Rev. Mex. Fis. \textbf{67}, 25--32 (2021), \href{https://doi.org/10.31349/RevMexFis.67.25}{doi:10.31349/RevMexFis.67.25}.

\bibitem{Romero2023}
R. Romero, ``Elko Spinors Revised,'' Rev. Mex. Fis. \textbf{69}, 020201 (2023), \href{https://doi.org/10.31349/RevMexFis.69.020201}{doi:10.31349/RevMexFis.69.020201}, arXiv:2207.08334 [hep-th].

\bibitem{WickWightmanWigner1952}
G. C. Wick, A. S. Wightman, and E. P. Wigner, ``The Intrinsic Parity of Elementary Particles,'' Phys. Rev. \textbf{88}, 101--105 (1952), \href{https://doi.org/10.1103/PhysRev.88.101}{doi:10.1103/PhysRev.88.101}.

\bibitem{SiglRaffelt1993}
G. Sigl and G. Raffelt, ``General Kinetic Description of Relativistic Mixed Neutrinos,'' Nucl. Phys. B \textbf{406}, 423--451 (1993), \href{https://doi.org/10.1016/0550-3213(93)90175-O}{doi:10.1016/0550-3213(93)90175-O}.

\bibitem{Volpe2013}
C. Volpe, D. V\"a\"an\"anen, and C. Espinoza, ``Extended Evolution Equations for Neutrino Propagation in Astrophysical and Cosmological Environments,'' Phys. Rev. D \textbf{87}, 113010 (2013), \href{https://doi.org/10.1103/PhysRevD.87.113010}{doi:10.1103/PhysRevD.87.113010}, arXiv:1302.2374 [hep-ph].

\bibitem{SerreauVolpe2014}
J. Serreau and C. Volpe, ``Neutrino--Antineutrino Correlations in Dense Anisotropic Media,'' Phys. Rev. D \textbf{90}, 125040 (2014), \href{https://doi.org/10.1103/PhysRevD.90.125040}{doi:10.1103/PhysRevD.90.125040}, arXiv:1409.3591 [hep-ph].

\bibitem{Kartavtsev2016}
A. Dobrynina, A. Kartavtsev, and G. Raffelt, ``Helicity oscillations of Dirac and Majorana neutrinos,'' Phys. Rev. D \textbf{93}, 125030 (2016), \href{https://doi.org/10.1103/PhysRevD.93.125030}{doi:10.1103/PhysRevD.93.125030}, arXiv:1605.04512 [hep-ph].

\bibitem{Vlasenko2014}
A. Vlasenko, G. M. Fuller, and V. Cirigliano, ``Neutrino Quantum Kinetics,'' Phys. Rev. D \textbf{89}, 105004 (2014), \href{https://doi.org/10.1103/PhysRevD.89.105004}{doi:10.1103/PhysRevD.89.105004}, arXiv:1309.2628 [hep-ph].

\bibitem{Cirigliano2015}
V. Cirigliano, G. M. Fuller, and A. Vlasenko, ``A New Spin on Neutrino Quantum Kinetics,'' Phys. Lett. B \textbf{747}, 27--35 (2015), \href{https://doi.org/10.1016/j.physletb.2015.04.066}{doi:10.1016/j.physletb.2015.04.066}, arXiv:1406.5558 [hep-ph].

\bibitem{Tian2017}
J. Y. Tian, A. V. Patwardhan, and G. M. Fuller, ``Prospects for Neutrino Spin Coherence in Supernovae,'' Phys. Rev. D \textbf{95}, 063004 (2017), \href{https://doi.org/10.1103/PhysRevD.95.063004}{doi:10.1103/PhysRevD.95.063004}, arXiv:1610.08586 [astro-ph.HE].

\bibitem{KobayashiLim2001}
M. Kobayashi and C. S. Lim, ``Pseudo Dirac Scenario for Neutrino Oscillations,'' Phys. Rev. D \textbf{64}, 013003 (2001), \href{https://doi.org/10.1103/PhysRevD.64.013003}{doi:10.1103/PhysRevD.64.013003}, arXiv:hep-ph/0012266.

\bibitem{Beacom2004}
J. F. Beacom, N. F. Bell, D. Hooper, J. G. Learned, S. Pakvasa, and T. J. Weiler, ``Pseudo-Dirac Neutrinos: A Challenge for Neutrino Telescopes,'' Phys. Rev. Lett. \textbf{92}, 011101 (2004), \href{https://doi.org/10.1103/PhysRevLett.92.011101}{doi:10.1103/PhysRevLett.92.011101}, arXiv:hep-ph/0307151.

\bibitem{FiorilloRaffeltSigl2024}
D. F. G. Fiorillo, G. G. Raffelt, and G. Sigl, ``Collective Neutrino--Antineutrino Oscillations in Dense Neutrino Environments?,'' Phys. Rev. D \textbf{109}, 043031 (2024), \href{https://doi.org/10.1103/PhysRevD.109.043031}{doi:10.1103/PhysRevD.109.043031}, arXiv:2401.02478 [hep-ph].

\bibitem{Balantekin2019}
A. B. Balantekin, A. de Gouv\^ea, and B. Kayser, ``Addressing the Majorana vs. Dirac Question with Neutrino Decays,'' Phys. Lett. B \textbf{789}, 488--495 (2019), \href{https://doi.org/10.1016/j.physletb.2018.11.068}{doi:10.1016/j.physletb.2018.11.068}, arXiv:1808.10518 [hep-ph].

\end{thebibliography}
\end{document}